\documentclass[twocolumn, dvipnames]{aastex7}
\usepackage[autostyle, english = american]{csquotes}
\MakeOuterQuote{"}
\usepackage{hyperref}
\usepackage[dvipsnames]{xcolor}
\hypersetup{
    colorlinks=true,
    linkcolor=Orange,
    filecolor=Orange,      
    urlcolor=Orange,
    citecolor=Orange,
    }
\usepackage[T1]{fontenc}
\usepackage{graphicx}	
\usepackage{amsmath}	
\usepackage{listings}
\newcommand{\flc}{\texttt{flc}}

\defcitealias{gaiadr3}{Gaia Collaboration et al. 2023}
\defcitealias{gaia1}{Gaia Collaboration et al. 2016}

\begin{document}

\shorttitle{Proper Motion of Draco II}
\shortauthors{J. T. Warfield et al.}
\title{The Proper Motion of Draco II with HST using Multiple Reference Frames and Methodologies}



\author[0000-0003-1634-4644]{Jack T. Warfield}
\thanks{Jefferson Scholars Foundation Graduate Fellow}
\affiliation{Department of Astronomy,
The University of Virginia,
530 McCormick Road,
Charlottesville, VA 22904, USA
}
\email[show]{jtw5zc@virginia.edu}

\author[0000-0001-7494-5910]{Kevin A. McKinnon}
\affiliation{David A. Dunlap Department of Astronomy \& Astrophysics, University of Toronto, 50 St George Street, Toronto ON M5S 3H4, Canada
}
\email{kevin.mckinnon@utoronto.ca}

\author[0000-0001-8368-0221, sname='Sohn']{Sangmo Tony Sohn}
\affiliation{Space Telescope Science Institute, 3700 San Martin Drive, Baltimore, MD 21218, USA}
\email{tsohn@stsci.edu}

\author[0000-0002-3204-1742]{Nitya Kallivayalil}
\affiliation{Department of Astronomy,
The University of Virginia,
530 McCormick Road,
Charlottesville, VA 22904, USA
}
\email{njk3r@virginia.edu}

\author[0000-0002-1445-4877]{Alessandro Savino}
\affiliation{Department of Astronomy,
University of California,
Berkeley, CA, 94720, USA}
\email{asavino@berkeley.edu}

\author[0000-0001-7827-7825, sname='van der Marel']{Roeland P. van der Marel}
\affiliation{Space Telescope Science Institute, 3700 San Martin Drive, Baltimore, MD 21218, USA}
\affiliation{Center for Astrophysical Sciences, The William H. Miller III Department of Physics \& Astronomy, Johns Hopkins University, Baltimore, MD 21218, USA}
\email{marel@stsci.edu}

\author[0000-0002-6021-8760]{Andrew B. Pace}
\thanks{Galaxy Evolution and Cosmology (GECO) Fellow}
\affiliation{Department of Astronomy,
The University of Virginia,
530 McCormick Road,
Charlottesville, VA 22904, USA
}
\email{apace@virginia.edu}

\author[0000-0001-9061-1697]{Christopher T. Garling}
\affiliation{Department of Astronomy,
The University of Virginia,
530 McCormick Road,
Charlottesville, VA 22904, USA
}
\email{txa5ge@virginia.edu}

\author[0009-0002-1233-2013]{Niusha Ahvazi}
\thanks{Galaxy Evolution and Cosmology (GECO) Fellow}
\affiliation{Department of Astronomy,
The University of Virginia,
530 McCormick Road,
Charlottesville, VA 22904, USA
}
\email{nahvazi@virginia.edu}

\author[0000-0001-8354-7279]{Paul Bennet}
\affiliation{Space Telescope Science Institute, 3700 San Martin Drive, Baltimore, MD 21218, USA}
\email{pbennet@stsci.edu}

\author[0000-0002-2970-7435]{Roger E. Cohen}
\affiliation{Department of Physics and Astronomy, Rutgers the State University of New Jersey, 136 Frelinghuysen Rd., Piscataway, NJ 08854, USA}
\email{rc1273@physics.rutgers.edu}

\author[0000-0001-6464-3257]{Matteo Correnti}
\affiliation{INAF Osservatorio Astronomico di Roma, Via Frascati 33, 00078, Monteporzio Catone, Rome, Italy}
\affiliation{ASI-Space Science Data Center, Via del Politecnico, I-00133, Rome, Italy}
\email{matteo.correnti@inaf.it}

\author[0000-0003-4207-3788]
{Mark A.\ Fardal}
\affiliation{Eureka Scientific, 2452 Delmer Street, Suite 100, Oakland, CA 94602, U.S.A.}
\email{mfardal@eurekasci.com}

\author[0000-0001-5538-2614]{Kristen.~B.~W. McQuinn}
\affiliation{Space Telescope Science Institute, 3700 San Martin Drive, Baltimore, MD 21218, USA}
\affiliation{Department of Physics and Astronomy, Rutgers the State University of New Jersey, 136 Frelinghuysen Rd., Piscataway, NJ 08854, USA}
\email{kmcquinn@stsci.edu}

\author[0000-0002-8092-2077]{Max J. B. Newman}
\affiliation{Space Telescope Science Institute, 3700 San Martin Drive, Baltimore, MD 21218, USA}
\email{mnewman@stsci.edu}

\author[0000-0002-2732-9717]{Eduardo Vitral}\thanks{Royal Society Newton International Fellow}
\affiliation{Institute for Astronomy, University of Edinburgh, Royal Observatory, Blackford Hill, Edinburgh EH9 3HJ, UK2}
\email{eduardo.vitral@roe.ac.uk}



\begin{abstract}

    We present proper motion (PM) measurements for Draco II, an ultra-faint dwarf satellite of the Milky Way. These PMs are measured using two epochs of Hubble Space Telescope Advanced Camera for Surveys (HST/ACS) imaging separated by a 7~year time baseline.
    Measuring PMs of low-luminosity systems is difficult due to the low number of member stars, requiring a precise inertial reference frame.
    We construct reference frames using three different sets of external sources: 1) stars with Gaia DR3 data, 2) stationary background galaxies, and 3) a combination of the two.
    We show that all three reference frames give consistent PM results. We find that for this sparse, low-luminosity regime including background galaxies into the reference frame improves our measurement by up to $\sim2\times$ versus using only Gaia astrometric data.
    Using 301 background galaxies as a reference frame, we find that Draco II's systemic PM is $(\mu_{\alpha}^*, \mu_{\delta}) = (1.043\pm0.029, 0.879\pm0.028)$~mas/yr, which is the most precise measurement of the three we present in this paper.

\end{abstract}

\keywords{Dwarf spheroidal galaxies (420); Local Group (929); Proper motions (1295)}



\section{Introduction} \label{sec:intro}

Measuring the 3-D motions of galaxies in the Local Group (LG) is critical to understanding its dynamics and interaction history. Combining 3-D velocity vectors with the positions and distances of LG galaxies enables robust modeling of satellite orbits, which have proved invaluable to a number of analyses in galaxy evolution and cosmology. Satellite orbit models have revealed that several low-mass dwarf satellites fell into the halo of the Milky Way (MW) with the LMC \citep{Patel2020, patel+2024}, as expected in a cold dark matter cosmology. Orbital models reveal the interaction history of dwarf satellites with their hosts, which is critical for understanding environmental processes like tidal and ram pressure stripping that affect the dark matter, stars, and gas in satellite galaxies after infall (e.g., \citealt{dinescu2004}, \citealt{besla2007}, \citealt{sohn+2013}, \citealt{sohn+2020}, \citealt{Battaglia+2021}, \citealt{richstein+2022}, \citealt{bennet2024}). As the satellite orbits are sensitive to the gravitational potential of their hosts, satellite orbits have also been used to probe the mass of the MW and M31 dark matter halos (e.g., \citealt{zaritsky+1989}, \citealt{watkins+2010}, \citealt{vandermarel+2012}, \citealt{diaz+2014}, \citealt{patel+2017}, \citealt{patel+2023}).

The motions of galaxies (or stars) relative to Earth (or the Sun) are generally split into two principal components. One is the velocity of a galaxy towards or away from us, referred to as the radial or line-of-sight velocity. The line-of-sight velocity is typically measured spectroscopically. The other component is a galaxy's 2-D angular velocity in the plane of the sky, referred to as a galaxy's proper motion (PM). The PM is further split into the velocities in the directions of right ascension (RA), $\mu_{\alpha}^*$,\footnote{$\mu_{\alpha}^* = \mu_{\alpha} \cos{(\delta)}$, which accounts for the converging lines of longitude towards the poles and keeps $\mu_{\alpha}^*$ in "square" angular units.} and declination (Dec), $\mu_{\delta}$.

While PMs can be calculated from time-series analysis of a target's position, the magnitudes of these angular velocities are quite small. Even Barnard's Star---the star in the MW with the largest angular PM---moves at only $\sim$10.4 arcseconds per year (\citetalias{gaiadr3}). LG dwarfs, owing to their immense distances and tangential velocities of $\lesssim500$~km/s, have PMs orders of magnitude smaller than this, $\lesssim$~1.0 milliarcseconds per year (e.g., \citealt{sohn+2017}, \citealt{Longeard+2018}, \citealt{McConnachie2020}, \citealt{li+2021}, \citealt{pace+2022}, \citealt{Battaglia+2021}), and galaxies at the distance of M31 have PMs $\lesssim$~0.1 milliarcseconds per year (e.g., \citealt{sohn+2020}, \citealt{warfield+2023a}, \citealt{bennet2024}, \citealt{casettidinescu2024_and3}, \citealt{bennet+2025}).
PMs, therefore, have historically been the most challenging velocity component to measure, requiring either extensive time baselines between observations or extremely precise astrometry, or both.\footnote{For instance, \cite{halley1717} is often credited with making the first note of the PMs of stars in the MW after comparing his observations with those of classical antiquity era astronomers Ptolemy and Hipparchus. Still, discerning these motions was only possible due to the $\sim$1800~year baseline between observations. \cite{lequeux2014} shows the astrometric precision of the best eighteenth century catalogs to be $\sim$1 arcminute.} 

It has been only within the past half century that we have been able to discern the PMs of a large sample of MW stars (i.e., $N_{\rm stars} \gtrsim$~100,000). The Hipparcos (\citealt{perryman+1997}; operating 1989--1993) and Gaia (\citetalias{gaia1}; over 1.4~billion PMs, operating 2013--2025) space missions are widely credited for revolutionizing robust PM studies (though, predecessors such as the Catalog of Positions and Proper Motions, or PPM, have also made important contributions; \citealt{ppmcat}, \citealt{ppmxl}). 

PMs can be calculated by measuring a target's changing position versus an external reference frame given two or more epochs of imaging separated by the time baseline $\Delta T$.
Hipparcos and Gaia measured PMs by imaging stars over their mission lifespans via a scanning astrometry strategy \citep{lindegrenbastian2010}, then fitting these data to astrometric solutions that take into account the telescope's positioning, pointing, and movement relative to a global stationary reference frame defined by the positions of quasi-stellar objects (QSOs; described for Gaia and Hipparcos in \citealt{lindegren+2012}, and updated for Gaia~DR3 in \citealt{Lindegren21a}). However, the limited precision and depth of the observations from these missions have restricted their ability to measure the PMs of only the brightest stars in MW satellites.
Alternatively, external reference frames can be constructed with two-epochs of high-precision, deep exposure, space-based imaging from observatories (such as HST) using QSOs (e.g., \citealt{kallivayalil+2006}), background galaxies \citep{sohn+2012}, and the Gaia DR3 reference frame \citep{delPino+2022, warfield+2023a, mckinnon+2024}.

Each of these PM measurement methods have their own set of advantages and shortfalls. QSOs, where available, are still perhaps the best sources for PM measurements since their positions (as point sources) can be precisely determined and their locations on the sky remain approximately stationary, even over long time baselines, owing to their large distances. However, QSOs are fairly rare, and a single HST/ACS pointing may have no verified QSOs at all, as is the case in our Draco~II field.
Background galaxies, like QSOs, are stationary on the sky, and hundreds are present in the background of any single HST pointing given sufficient exposure times. However, the precision of centroiding any individual galaxy is made difficult by their extended (non-point-source) profiles, and the need for very deep imaging to get adequate signal-to-noise to model their profiles. Stars with Gaia measurements are found in much larger numbers than QSOs and their positions can be measured to high precision via point-spread function (PSF) fitting. However, the uncertainty of this reference frame will mostly be determined by 1.) the uncertainties on each individual star's Gaia PM measurement and 2.) systematics between the HST/ACS sky frame versus the Gaia sky frame. Additionally, for galaxies with many stars whose PMs can be measured directly by Gaia, the restriction of the size of the ACS field makes it often difficult to improve much on the uncertainties of the Gaia-only measurement, despite the added astrometric power of HST.
Taken all together, these considerations suggest the utility of making consensus PM determinations using a variety of methods.

Draco~II is an ulta-faint dwarf (UFD)\footnote{Since its discovery \citep{dr2discover}, the classification of Draco~II as a UFD has been debated \citep{2016MNRAS.458L..59M, Longeard+2018}. While mass-segregation measurements suggest classification as a star cluster \citep{2022MNRAS.510.3531B}, more recent measurements of Draco~II's metallicity spread \citep{2023ApJ...958..167F} and SFH \citep{2025arXiv250518252D} have bolstered its classification as a UFD.} satellite of the MW with $M_V\simeq -0.8$ and a distance of $\sim$22~kpc from the Sun \citep{Longeard+2018}. In this paper, we measure the PM of Draco~II using two epochs of HST/ACS imaging with an $\sim$7~year baseline using three semi-independent methods: constructing a reference frame from Gaia stars using \texttt{HubPUG}\footnote{\url{https://github.com/jackwarfield/HubPUG}} (\citealt{warfield+2023a}; \S\ref{sec:method1}), constructing a reference frame from background galaxies by adapting the \texttt{HubPUG} machinery (\S\ref{ssec:gals}), and constructing a reference frame from Gaia stars and background galaxies adapting the methodology of \texttt{BP3M}\footnote{\url{https://github.com/KevinMcK95/BayesianPMs}} (\citealt{mckinnon+2024}; \S\ref{sec:bp3m}).
We present an internal comparison of the three methods, an external comparison to PMs measured from the Gaia DR3 catalog \citep{Battaglia+2021, pace+2022}, and a discussion of systematics between measurements in \S\ref{sec:results}. In \S\ref{sec:conclusions}, we discuss our conclusions.

\section{Observations and Data} \label{sec:data}

Draco~II's first-epoch observations were taken in 2017 March with HST/ACS's F814W filter (Wide-\textit{I}, $\sim$700--950 nm), resulting in four \flc\ images\footnote{The \flc\ filetype indicates calibrated, charge-transfer efficiency (CTE) corrected, and flat-fielded single-exposure images.} with exposure times of $\sim$1150 seconds each (HST-GO 14734, PI N. Kallivayalil).
Second-epoch observations were taken in 2024 March (HST-GO 17473, PI J. T. Warfield), where Draco~II was observed with the same filter, pointing, and orientation as the first-epoch observations in order to maximize field overlap and to minimize possible systematic effects from geometric distortion and CTE (e.g., as suggested by \citealt{sohn+2012}).
All the HST data used in this paper can be found in MAST: \dataset[10.17909/240g-c756]{http://dx.doi.org/10.17909/240g-c756}.
These images were retrieved from the MAST archive, and catalogs of sources in each image were generated by running \texttt{hst1pass}\footnote{\url{https://www.stsci.edu/\textasciitilde jayander/HST1PASS/}} \citep{hst1pass}---a \texttt{Fortran}-based effective point-spread function (ePSF) fitting routine for HST ACS and Wide Field Camera 3 images. \texttt{hst1pass} performs this ePSF-fitting on \flc\ files at the sub-pixel level, making it ideal for tasks like PM measurements that require very precise positions measurements. After finding positions for stars in pixel coordinates for each \flc, we then apply the ACS pipeline geometric-distortion correction (GDC; \citealt{gdc1}, \citealt{gdc2}) to each position.

In order to generate color-magnitude diagrams (CMDs) for identifying likely Draco~II member stars in our catalogs, we also generated a photometric catalog from both the F814W first-epoch HST observations as well as observations in the F606W filter, taken in the same epoch. This catalog was generated using \texttt{DOLPHOT} \citep{dolphin2000b, dolphin2016}, which is software for HST (and JWST) drizzled-image photometry. The photometric accuracy of \texttt{DOLPHOT} over \texttt{hst1pass} is not strictly required for our purposes, but we opted to make use of this catalog to take advantage of the quality parameters for cutting out contaminant (non-stellar) sources. We ran \texttt{DOLPHOT} in the manner described by \cite{Williams2014} and \cite{savino+2022, savino+2023}, including cutting the catalog down to likely stellar sources by applying the cuts (to both sets of per-filter values): ${\rm S/N} \geq 4$, ${\rm Sharp}^2 \leq 0.2$, ${\rm Crowd} \leq 0.75$, and ${\rm Round} \leq 3$.\footnote{Definitions for these parameters can be found on the \texttt{DOLPHOT} user's manual webpage: \url{https://dolphot-jwst.readthedocs.io/en/latest/post-processing/catalogs.html}.} This catalog was then cross-matched with the catalogs from \texttt{hst1pass}.

\section{Method 1: HubPUG} \label{sec:method1}

\texttt{HubPUG} (Hubble Proper motions Utilizing Gaia; \citealt{warfield+2023a}) is a software tool that calculates the PM of resolved stellar systems that have been observed with two epochs of HST imaging. To use this tool, we only need to provide source catalogs for each individual \flc\ image; magnitude and PSF fit-quality ($q_{\rm fit}$) limits for the range of stars in the target galaxy that we would like to be used; magnitude and $q_{\rm fit}$ limits for the range of sources to be cross-matched against Gaia; and tolerance thresholds for transforming each catalog to a common reference frame.
With these, \texttt{HubPUG} first iteratively transforms each \flc\ source catalog to the stationary, comoving frame of the target galaxy (i.e., the frame where well-measured stars in Draco~II are stationary across epochs, on-average). The transformation script is seeded with the catalog of stars determined likely to be Draco~II members based on by-eye CMD cuts (this is the same list of stars used in the corrections discussed in \S\ref{ssec:gals} and shown in Figure~\ref{fig:loc_offsets}). 
Following this coordinate transformation, bright sources in each \flc\ catalog are matched to the Gaia~DR3 catalog, and then each \flc\ catalog is crossmatched among the others. After this crossmatch, the average position of each star in all first-epoch catalogs versus all second-epoch catalogs for each Gaia star is recorded.
We then subtract off each Gaia source's individual PM (as reported in Gaia DR3), so that each Gaia source now independently indicates the reflex, systemic PM  of the comoving frame, which is the absolute systemic PM of the target galaxy.

\begin{figure*}
    \plotone{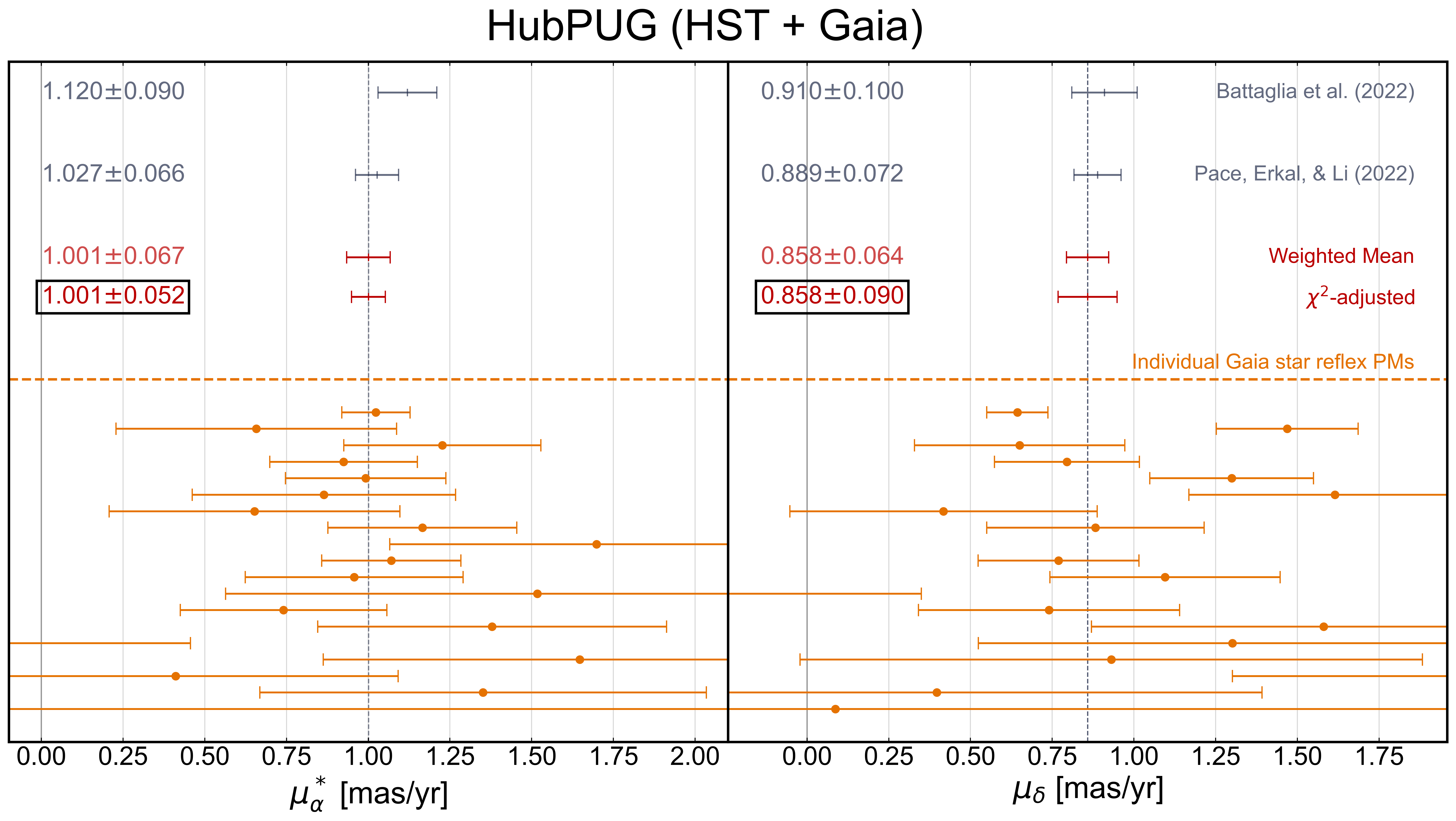}
    \caption{Summary of the results for the PM measurement of Draco~II from \texttt{HubPUG} utilizing the Gaia frame-of-reference. The left-hand and right-hand panels show the results for $\mu_\alpha^*$ and $\mu_\delta$, respectively. The orange points below the dashed orange line represent each of the PM estimates of Draco II that come from individual Gaia stars, sorted by magnitude (brightest at the top). The red points (third and fourth from the top) are the inverse-variance weighted mean of the individual PMs, with the dotted line tracing the median throughout the entire plot, and showing the unmodified weighted mean uncertainties as well as the uncertainties rescaled by the $\chi^2_\nu$. The top two blue points are the PM for Draco~II as reported by \protect\cite{Battaglia+2021} and \protect\cite{pace+2022}, both calculated using only data from Gaia eDR3. \label{fig:hpstars}}
\end{figure*}
We then calculate the systemic PM of Draco~II as the inverse-variance weighted mean of the reflex PMs of the Gaia stars. As we will touch on in \S\ref{sec:results}, we rescale uncertainties on our measurement by a factor of $\sqrt{\chi^2_\nu}$. A summary plot of our measurement of Draco~II's systemic motion using \texttt{HubPUG} is shown in Figure~\ref{fig:hpstars}.

\section{Method 2: Measurement from Background Galaxies} \label{ssec:gals}

In addition to using Gaia DR3 as a reference frame, \texttt{HubPUG} allows matching sources against a catalog of known QSOs (the Million Quasars Catalogue; \citealt{milliquas2021, flesch2023}). However, there are no QSOs present in the ACS field-of-view for Draco~II.

Similar to QSOs, background galaxies (distant, unresolved galaxies in the field) are a useful reference frame because their positions on the sky are roughly stationary relative to each other on human timescales. Therefore, we re-purposed the module of \texttt{HubPUG} meant for measuring PMs using QSOs for use with background galaxies by generating catalogs of background galaxy positions as a substitute for QSO positions.

The relevant deviation from, and difficulty added, versus the QSO case stems from our ability to find positions for (i.e., centroid) background galaxies in a consistent manner across multiple images and epochs. Because QSOs are point sources, we can centroid them using normal stellar PSF methods. Background galaxies have extended profiles. Not only do these profiles differ galaxy-to-galaxy, but their centroids are also less stable than those of point sources and can exhibit time-dependent filter-, geometric distortion-, and detector-based systematics.

\begin{figure}
    \plotone{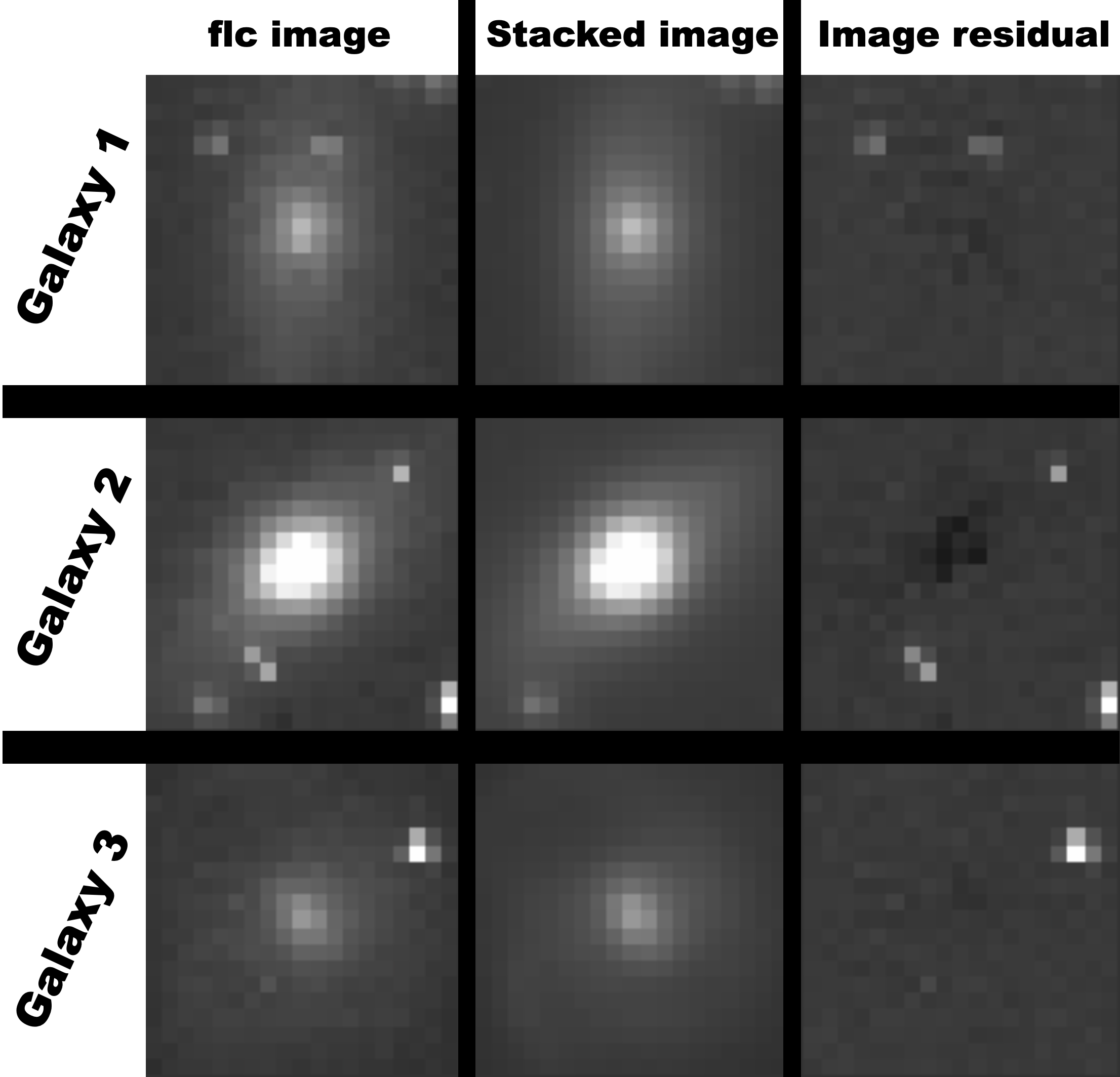}
    \caption{Three examples of background galaxies in the Draco~II field, with each row being a distinct galaxy. Left column: the 11$\times$11 pixel stamp of the galaxy in the second-epoch image \texttt{jf9q02tvq}. Middle column: the convolved galaxy profile template built from the first-epoch image stack. Right column: difference image of the first column minus the second column. \label{fig:gsf_example}}
\end{figure}
To overcome this issue, we adopt a method similar to that proposed by \cite{mahmudanderson2008} and first applied by \cite{sohn+2012, sohn+2013}, using a template-fitting approach to find the positions of background galaxies in each image. The steps are as follows:
\begin{enumerate}
    \item Using CMD cuts, cut down the per-\flc\ image source catalogs to stars that are probable members of Draco~II;
    \item Cross-match the per-\flc\ catalogs for the four first-epoch images;
    \item Iteratively solve for the six-dimensional linear transformation\footnote{$X_{\rm trans} = A_X + B_X X + C_X Y$;\\ $Y_{\rm trans} = A_Y + B_Y X + C_Y Y$.} (6DLT) for each catalog to the median catalog until the scatter for a given star is $\leq0.05$~pixels, with each iteration cutting stars from the transformation that do not fall within this threshold;
    \item Use these transformations to generate a stacked, two-times super-sampled and geometric distortion-corrected image from the four first-epoch images in the same fashion as \S3.4 of \cite{sohn+2012}, and similar to the process of \texttt{Drizzle} \citep{drizzle2002} and \texttt{iDrizzle} \citep{drizzle2011};
    \item Cross-match the second-epoch per-\flc\ image source catalogs against the catalog of star positions in the first-epoch stacked image;
    \item Iteratively solve for the 6DLT between each second-epoch \flc\ and the stacked image;
    \item Create a reduced star catalog of the brightest (but not saturated), best measured stars for the stacked catalog and each second epoch \flc;
    \item For each \flc, build a 7$\times$7 image convolution kernel between the stacked image and each second-epoch distortion-corrected \flc's (to account for centroid shifts owed to changes in the PSF between epochs) by minimizing the average difference image between each star in the reduced catalog;
    \item Compile a catalog of background galaxies and their approximate positions;\footnote{For this work, we compiled this catalog by combing the images in \texttt{SAOImageDS9} \citep{ds9} at varying zoom levels and picking out sources by-eye. In principle, these catalogs could be constructed using automated tools, such as \texttt{Source-Extractor} \citep{sextractor} or \texttt{GALFIT} \citep{galfit}. However, we have found that relying only on these automated tools often leads to missing a number of useful background sources.} 
    \item Record the template profile of each galaxy---which we refer to as the "GSF"---as the 22$\times$22 pixel "stamp" (in super-sampled units; or 11$\times$11 in native \flc\ pixel space) around each background galaxy in the stacked image;
    \item Find the position of each galaxy in each \flc\ from both epochs by:
	\begin{enumerate}
	    \item Down-sampling the template to the \flc\ pixel scale and, if comparing to a second-epoch image, convolving the template by the associated 7$\times$7 image kernel;
	    \item Overlaying the GSF template on the distortion-corrected \flc\ image using the 6DLT transformation parameters found between the stacked frame and each \flc\ image;
	    \item Blotting the template with sub-pixel offsets versus the \flc\ image and subtracting the template image to find the offset with the minimum subtraction residuals (e.g., Figure~\ref{fig:gsf_example}), which defines that galaxy's position in that image;
	\end{enumerate}
\end{enumerate}
Following Step 11, we calculate the mean position of each background galaxy in the comoving frame across all first-epoch images and across all second-epoch images in which they appear. In other words, we are able to repeat the same process as what was done with Gaia stars in \S\ref{sec:method1}, but using background galaxies with individual PMs that are approximately zero as the independent references for the comoving frame's PM rather than Gaia stars.

\begin{figure*}
    \includegraphics[width=\textwidth]{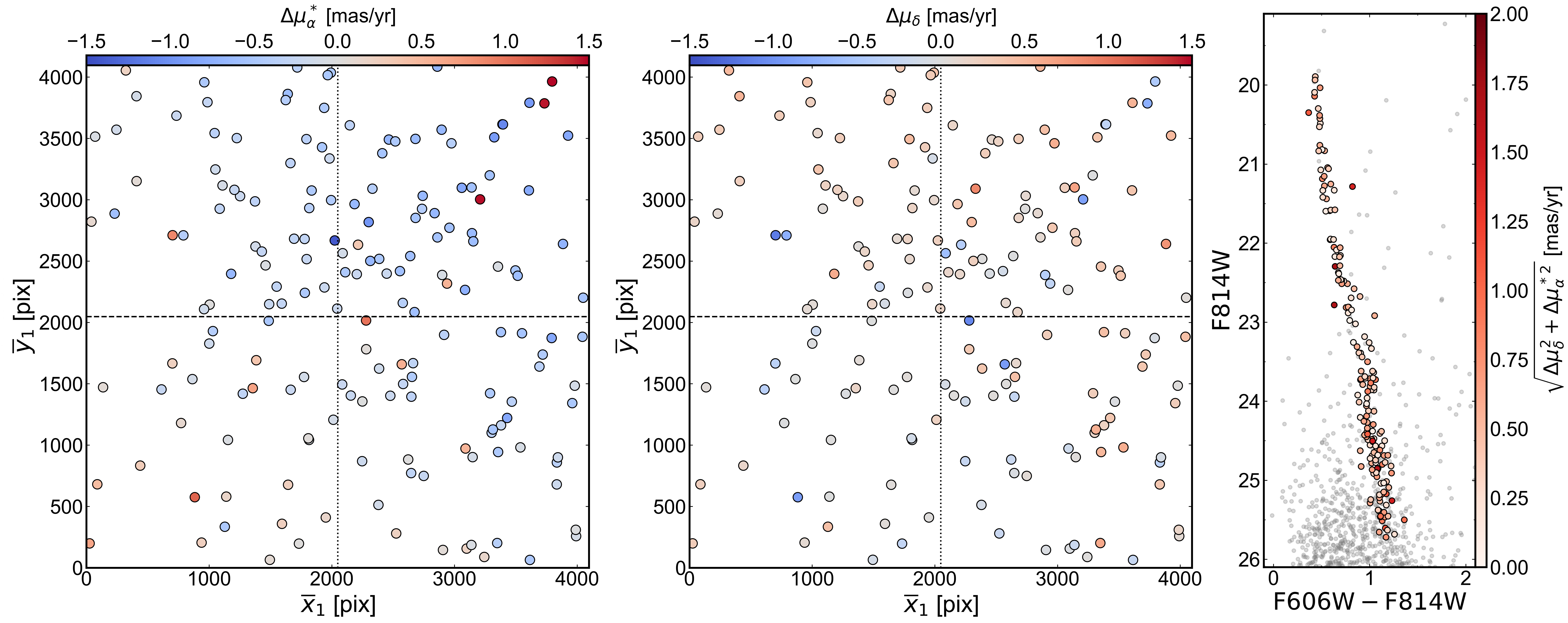}
    \caption{\textit{Left}: Average first-epoch \flc\ $(x,y)$ positions of Draco~II member stars used in the local and chip corrections, colored by the excess PM of each star along the right ascension axis. Middle: Average first-epoch \flc\ $(x,y)$ positions of Draco~II member stars used in the the local correction, colored by the excess PM of each star along the declination axis. \textit{Right}: CMD of the Draco~II member stars used in the the local correction, colored by the total excess PM of each star. This is the same set of stars used as the initial input for the transformation to the comoving frame. Additionally, in gray, we plot all objects in our \texttt{DOLPHOT} stellar catalog. \label{fig:loc_offsets}}
\end{figure*}
We additionally apply corrections to the reflex PM of each individual galaxy using the excess PMs measured from CMD-selected Draco~II member stars.
In Figure~\ref{fig:loc_offsets}, we plot each Draco~II star based on its average epoch 1 \flc\ position, colored by the residual PM in RA (left) and Dec (middle).
Since Draco~II stars should be stationary in the comoving frame, these residuals are products of non-linear, time-dependent offsets between the GDC-corrected HST frames of the two epochs.

There are several different possible approaches to decide which subset of stars should be used to derive these corrections. For instance, \cite{sohn+2020} suggests a "local" correction, using stars within a typical radius of 200~pixels around each background galaxy. However, for sparse systems like Draco~II, radii of this specific size are small enough that many galaxies with useful measurements cannot be used due to not being in the vicinity of an adequate number of Draco~II stars.
On the other end, \cite{casettidinescu+2025}, for example, calculate a similar correction using the average offset across each entire chip due to the nature of their data (a combination of ACS and archival WFPC2 observations). This approach is effective at correcting offsets that are partially products of the discontinuous nature of the GDC solution between the independent chips that make up the instrument's detector, and allows every galaxy with a measured position to be used.

To see the difference that this makes, we test these two implementations. The first is to perform the "local" correction (LC). We set the radius for this LC by first calculating the average number density of CMD-selected stars per $4096 \times 2048$~pixel$^2$ ACS chip that combine to make the full detector (99 stars on chip 1, 77 on chip 2). I.e., for chip 1, $n = \frac{99~{\rm stars}}{4096~{\rm pix} \times 2048~{\rm pix}}$. Because we want \emph{at least} three stars per-galaxy for the correction, we choose a radius that would give us, on average, $3\times2=6$ stars for any $(x,y)$ position, so $r = \sqrt{\frac{1}{\pi}\frac{6~{\rm stars}}{n}} \approx 402~{\rm pixels}$ for chip 1. With this correction, we find the PM of Draco~II, calculated as the 3-$\sigma$ clipped weighted-average of 234 background galaxies' reflex PMs (with $\chi^2$-adjusted uncertainties), to be:
\begin{equation}
    \begin{pmatrix} \mu_{\alpha}^* \\ \mu_{\delta}^{\phantom{*}} \end{pmatrix}_{\rm LC}
        =
    \begin{pmatrix} 1.042 \,\pm\,0.033 \\
        0.916 \,\pm\,0.031 \end{pmatrix}~{\rm mas/yr}.
\end{equation}

The second approach that we tested is to split the detector into four quadrants and calculate the correction for each background galaxy using every Draco~II star that falls within the same quadrant as that galaxy.
The motivation for splitting into quadrants rather than chips is that, although a portion of these excess motions may be from offsets between the actual GDC solution for each image, a majority of the offsets are likely due to CTE effects. Most effects of CTE on astrometry/photometry are corrected for in the ACS pipeline when creating the \flc\ image files. However, CTE affects fainter sources more significantly than brighter sources, with \cite{andersonryon2018} showing that sources with instrumental magnitudes fainter than $\sim$~6.5 in fields with low sky counts can lose up to 0.5~mags, owed to excess CTE effects not accounted for by the pipeline correction (this is also why this correction is not necessary in \S\ref{sec:method1}, where bright stars make up the reference frame).
The ACS detector reads out to four amplifiers total, each responsible for half of a single chip, and so the excess CTE for each source will be correlated with the star in the same quadrant as it.
Calculating the correction in this way, we are able to use 301 background galaxies, and we get values:
\begin{equation}
    \boxed{\begin{pmatrix} \mu_{\alpha}^* \\ \mu_{\delta}^{\phantom{*}} \end{pmatrix}_{\rm quad}
        =
    \begin{pmatrix} 1.043 \,\pm\,0.029 \\
        0.879 \,\pm\,0.028 \end{pmatrix}~{\rm mas/yr}.}
\end{equation}
\begin{figure*}
    \plotone{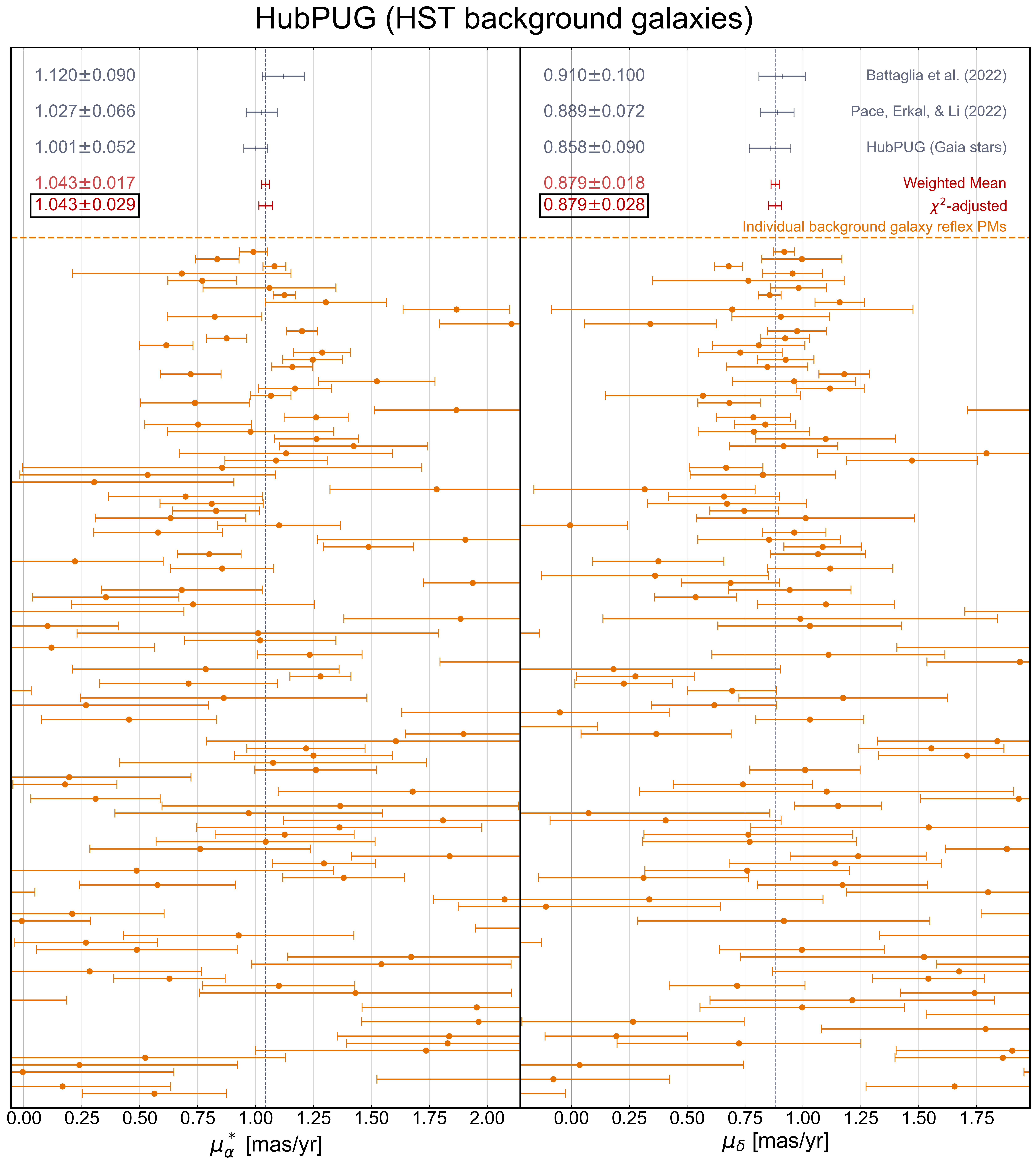}
    \caption{Summary of the results for the PM measurement of Draco~II from \texttt{HubPUG} utilizing background galaxies as a frame-of-reference. The left-hand and right-hand panels show the results for $\mu_\alpha^*$ and $\mu_\delta$, respectively. The orange points below the dashed orange line represent each of the PM estimates of Draco II that come from individual background galaxies, sorted by magnitude (brightest at the top). In order to maintain plot legibility, we only plot the 118 of the 301 background galaxies used in the calculation with total PM uncertainties $<1.0$~mas/yr. The red points (fourth and fifth from the top) are the inverse-variance weighted mean of the individual PMs, with the dotted line tracing the median throughout the entire plot, and showing the unmodified weighted mean uncertainties as well as the uncertainties rescaled by the $\chi^2_\nu$. The top two blue points are the PM for Draco~II as reported by \protect\cite{Battaglia+2021} and \protect\cite{pace+2022}, both calculated using only data from Gaia eDR3, and the third point is our HubPUG weighted-mean value presented in \S\ref{sec:method1}. \label{fig:hpgals}}
\end{figure*}
Because the difference between the two sets of random uncertainties corresponds directly with the difference in the number of background galaxies used in each result,\footnote{Using only the galaxies available for the LC implementation, but with the chip-based corrections gives the measurements $\mu_{\alpha}^* = 1.045 \pm 0.032$~mas/yr and $\mu_{\delta} = 0.913 \pm 0.030$~mas/yr.} and because the two results are in statistical agreement with each-other, we adopt the chip-based correction as our preferred measurement for this method. A summary plot for this result can be found in Figure~\ref{fig:hpgals}.

\section{Method 3: BP3M} \label{sec:bp3m}

We also measure PMs of Draco II stars using the Bayesian Positions, Parallaxes, and Proper Motions (\texttt{BP3M}) tool, which was originally described in \citet{mckinnon+2024}. At a high level, \texttt{BP3M} uses Gaia-measured astrometry of stars as prior information in combination with the positions of those stars in HST images. Then, \texttt{BP3M} simultaneously aligns all HST images onto the Gaia reference frame as well as measures improved astrometry for each star in a fully Bayesian fashion. This technique propagates the uncertainty of the HST image alignment to the final PMs, and it is guaranteed to return PMs that are at least as precise as the input prior Gaia PMs. 

The version of \texttt{BP3M} we use for the analysis in this work has been improved over the original version. Specifically, the new method re-writes the statistical arguments in \citet{mckinnon+2024} to define closed-form posterior distributions for the transformation solutions and stellar astrometry; this means that we no longer need to run slow MCMC steps. As a result, analyzing all Draco II HST images simultaneously takes less than $\sim20$ minutes using a single core on an Apple M3 chip. These improvements will be discussed in detail in an upcoming publication (McKinnon et al. in prep.).

In addition to using the Gaia stars identified in each HST image, the \texttt{BP3M} analysis also uses the background galaxies identified in Section~\ref{ssec:gals}. The galaxies are treated similarly to the stars, but instead of having Gaia-based priors on the PM and parallax, they have priors that ensure their parallax and proper motion are close to 0 (e.g. a Gaussian prior with mean 0 and width of $10^{-3}$ mas or mas/yr). Using these background galaxies is particularly useful for reducing the uncertainty in the transformation solution of each HST image, resulting in more precise and accurate PMs of the Draco II stars. 

The exact steps for the \texttt{BP3M} analysis is as follows:
\begin{enumerate}
    \item Run \texttt{GaiaHub} \citep{delPino+2022} on each HST image as a preprocessing step to automatically identify HST sources and cross-match to Gaia;
    \item Run \texttt{BP3M} on all HST images simultaneously using the cross-matches HST-Gaia stars and background galaxies;
    \item Cross-match faint HST-only stars (i.e. $G>21.5$~mag, where the majority of Draco II stars are found; see Figure~\ref{fig:loc_offsets}) between the different HST images;
    \item Use the final HST image transformation solution posterior distributions and the cross-matched HST-only sources to measure PMs for Draco II stars. 
\end{enumerate}
Expanding on this last step, we draw samples of the transformation parameters for all of the HST images from the full posterior distribution (i.e., $6\times N_{\mathrm{images}}$ parameters), retaining the full correlation between the parameters from each image. For a given draw of the transformation parameters, we transform the HST positions (and uncertainties) of the stars to Gaia-frame RA, Dec coordinates, again retaining the full correlation between the RA, Dec coordinates between the HST images. Measuring the stellar proper motion is then as simple as fitting a vector to the observed HST RA, Dec positions as a function of time for each star. We repeat this process for each of the posterior transformation solution draws to marginalize over the effect of the uncertainty in transformation solutions. In the end, take means and covariances of the PM samples for each star to define the final PM posterior distribution. 

Naturally, the PMs for each star are correlated with all of the other stars due to the impact of image alignment on relative positions on the sky. We ensure to take this correlation into account when measuring the bulk motion of Draco II using the \texttt{BP3M} PMs.

\begin{figure*}
    \plotone{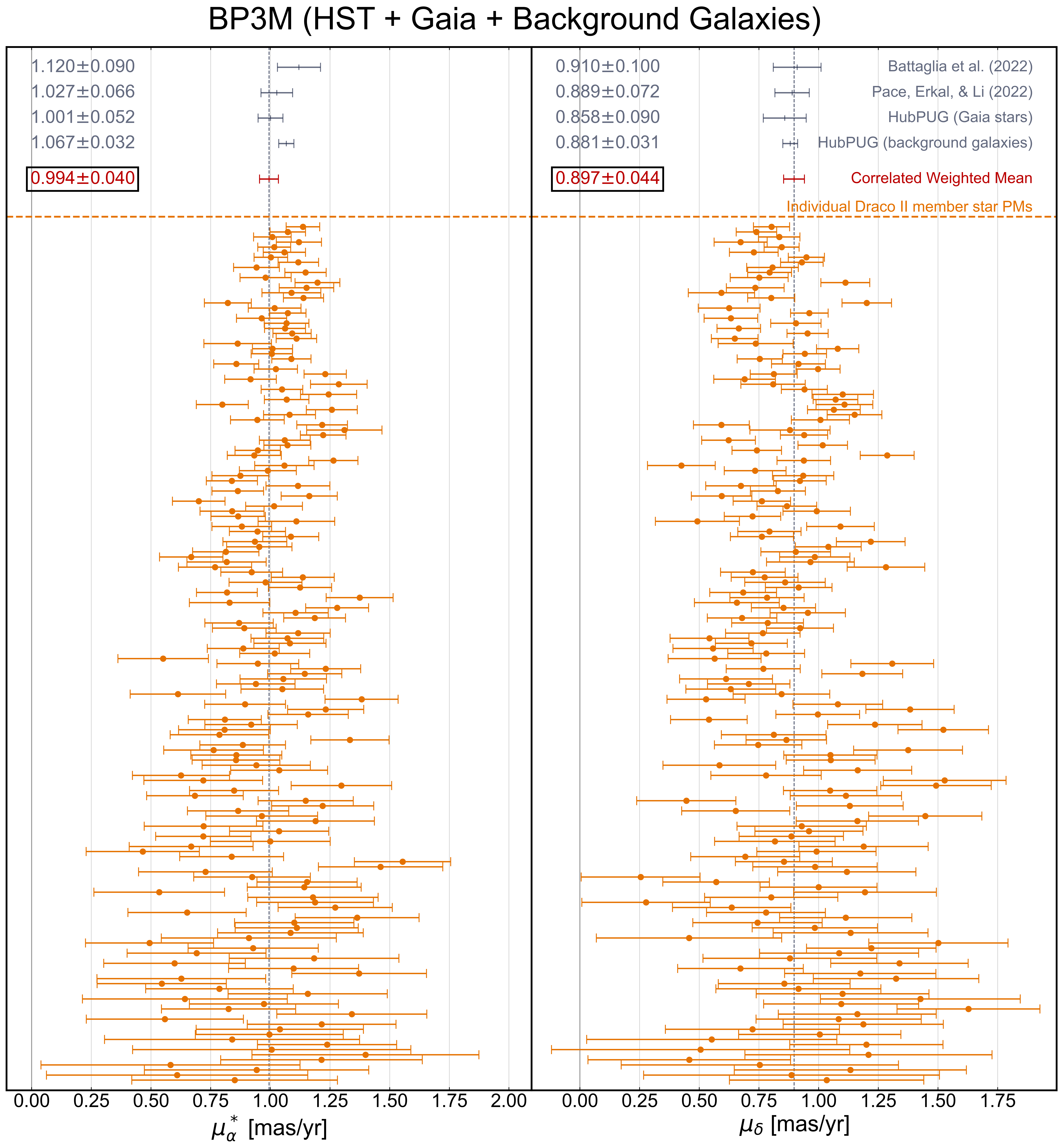}
    \caption{Summary of the results for the PM measurement of Draco~II from \texttt{BP3M} utilizing a combination of Gaia stars and background galaxies. The left-hand and right-hand panels show the results for $\mu_\alpha^*$ and $\mu_\delta$, respectively. The orange points below the dashed orange line represent each of the PM estimates of Draco II that come from individual stars that are members of Draco~II, sorted by magnitude (brightest at the top). The red point (fifth from the top) is the mean systemic motion of Draco~II taking into account the values from individual stars and the correlations between their PMs, with the dotted line tracing the median throughout the entire plot. The top two blue points are the PM for Draco~II as reported by \protect\cite{Battaglia+2021} and \protect\cite{pace+2022}, both calculated using only data from Gaia eDR3, and the third and fourth points are our HubPUG weighted-mean values presented in \S\ref{sec:method1} and \S\ref{ssec:gals}. \label{fig:bp3m}}
\end{figure*}

We next pass these PMs to \texttt{scikit-learn}'s Gaussian Mixture tool to identify Draco II likely members based on position in PM-space. We find that two components -- broadly corresponding to ``Draco II'' and ``background'' -- are able to cleanly separate the PMs. We keep stars with $> 0.5$ membership probability and then measure a bulk PM for the population. Because the Gaussian mixture probabilities did not take PM uncertainties into account, we do a final iteration of cleaning where we remove stars that have a $\chi^2 > 3.5^2$ when compared with the measured PM mean. This results in 169 Draco II stars used to measure a bulk PM of:
\begin{equation}
    \begin{pmatrix} \mu_{\alpha}^* \\ \mu_{\delta}^{\phantom{*}} \end{pmatrix}
        =
    \begin{pmatrix} 0.994 \,\pm\,0.040 \\
        0.897 \,\pm\,0.044 \end{pmatrix}~{\rm mas/yr}.
\end{equation}
A summary plot of our systemic PM measurement made using \texttt{BP3M} can be found in Figure~\ref{fig:bp3m}.

\section{Proper Motion Results} \label{sec:results}

\subsection{Summary} \label{ssec:summary}
\begin{figure*}
    \plotone{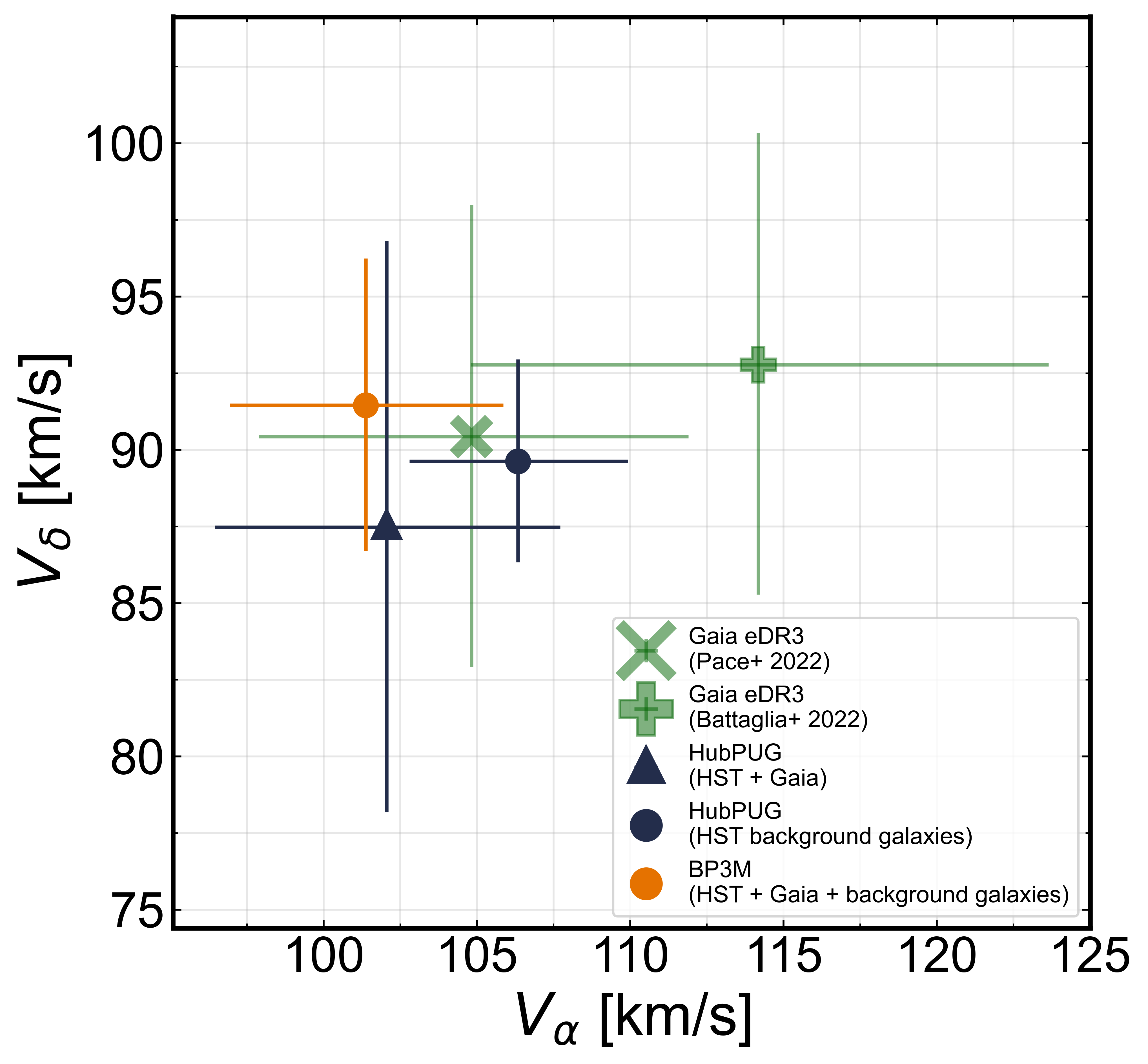}
    \caption{A comparison between the tangential velocity values measured for Draco~II from the three methodologies explored in this paper---using \texttt{HubPUG} with Gaia stars as a reference, using \texttt{HubPUG} with background galaxies as a reference, and using the \texttt{BP3M} methodology with two epochs of imaging data. The velocites calculated from the results using Gaia eDR3 data from \protect\cite{Battaglia+2021} and from \protect\cite{pace+2022} are also shown for comparison. We also plot the velocity calculated using the weighted-mean of the \texttt{HubPUG} background galaxies-based PM and the PM from \protect\cite{pace+2022}. Physical PM values are calculated assuming a distance of 21.5~$\pm$~0.4~kpc \protect\citep{Longeard+2018}. \label{fig:comparison}}
\end{figure*}
\begin{deluxetable*}{lcccccc}
    \tablecaption{PM Measurements for Draco~II.\label{tab:pmresults}}
    \tablehead{
	\colhead{Method} & \colhead{$\mu_{\alpha}^*$} & \colhead{$\mu_{\delta}$} & \colhead{$N$} & \colhead{$\chi_\nu^2$} & \colhead{$V_{\alpha}$} & \colhead{$V_{\delta}$} \\
	\colhead{} & \colhead{(mas/yr)} & \colhead{(mas/yr)} & \colhead{} & \colhead{($\alpha$, $\delta$)} & \colhead{(km/s)} & \colhead{(km/s)}
    }
    \startdata
    \textbf{\uline{This work}:} & & & &\\
    \texttt{HubPUG} (HST + Gaia) & $1.001\,\pm\,0.052$ & $0.858\,\pm\,0.090$ & $19$ & $(0.61, 1.95)$ & $102.08\,\pm\,5.7$ & $87.5\,\pm\,9.3$ \\
    \texttt{HubPUG} (HST BGs) & $1.043\,\pm\,0.029$ & $0.879\,\pm\,0.028$ & $301$ & $(2.97, 2.46)$ & $106.4\,\pm\,3.6$ & $89.6\,\pm\,3.3$ \\
    \texttt{BP3M} (HST + Gaia + BGs) & $0.994\,\pm\,0.040$ & $0.897\,\pm\,0.044$ & $169$ & $-$ & $101.4\,\pm\,4.5$ & $91.5\,\pm\,4.8$ \\ \\
    \textbf{\uline{Literature}:} & & & &\\
    Pace+22 (Gaia GMM)\tablenotemark{a} & $1.027\,\pm\,0.066$ & $0.887\,\pm\,0.072$ & $20$ & $-$ & $104.9\,\pm\,7.0$ & $90.5\,\pm\,7.5$ \\
    Battaglia+22 (Gaia GMM)\tablenotemark{b} & $1.12\,\pm\,0.09$ & $0.91\,\pm\,0.10$ & $29$ & $-$ & $114.2\,\pm\,9.4$ & $92.8\,\pm\,10.3$
    \enddata
    \tablecomments{BGs~$=$~background galaxies, GMM~$=$~Gaussian mixture model. $N$ is the number of sources that were used as independent measures of the PM (Gaia stars or background galaxies for \texttt{HubPUG}, Draco II member stars for \texttt{BP3M}). For literature Gaia eDR3 measurements, $N$ is the number of Gaia stars tagged as Draco II members. For values calculated using an inverse-variance weighted mean, $\chi_\nu^2$ values that were used to adjust the uncertainties to what is reported in the table are given. Velocities in km/s are calculated assuming a distance of $21.5\pm0.4$~kpc \protect\citep{Longeard+2018}, and are calculated using random draws from the PM and distance distributions.}
    \tablenotetext{a}{\cite{pace+2022}}
    \tablenotetext{b}{\cite{Battaglia+2021}}
\end{deluxetable*}
Our three PM measurements for Draco~II are summarized in Table~\ref{tab:pmresults}, as well as two representative PM results using Gaia~DR3 astrometric data in Gaussian mixture models (GMMs) from \cite{pace+2022} and \cite{Battaglia+2021}.\footnote{PM values are equivalent between Gaia eDR3 and Gaia DR3.}
We also give tangential velocity measurements expressed in physical units of km/s, adopting a heliocentric distance of $21.5\pm0.4$~kpc to Draco~II \citep{Longeard+2018}. The uncertainty in this distance measurement is included in the uncertainty on our tangential velocity values. In Figure~\ref{fig:comparison}, we plot the these tangential velocity measurements against each other.

\subsection{Zero-point Systematic Uncertainty}
Although we are able to report the estimated statistical uncertainty of each measurement in Table~\ref{tab:pmresults}, it is not entirely clear to what degree systematic uncertainties---including those associated with the chosen methodologies themselves---affect each measurement.

Zooming in on the \texttt{HubPUG} measurements, each individual Gaia star (or background galaxy) has uncertainties constructed as (in units of mas/yr):
\begin{equation}
    \sigma_i^2 = (\delta X_1^2 + \delta X_2^2 + \sigma_{\rm trans}^2 + \sigma_{\rm local}^2 + {\tt PM\_ERR}^2),
\end{equation}
where $\delta X_1$, $\delta X_2$ are the scatter on the position of an individual star/galaxy in the first and second epoch, respectively; $\sigma_{\rm trans}$ is the residual scatter of Draco~II member stars from the mean after transforming each \flc\ frame; $\sigma_{\rm local}$ is the scatter in the local or quadrant correction, applied to each background galaxy ($=0$ for Gaia stars); and \texttt{PM\_ERR} is the uncertainty on the Gaia \texttt{PMRA} or \texttt{PMDEC} of an individual star ($=0$ for galaxies).
For the Gaia star-only \texttt{HubPUG} measurement, \texttt{PM\_ERR} is the dominant source of uncertainty, and so those values, the number of Gaia stars used, and the time baseline establish a floor to the possible precision.
However, \texttt{HubPUG} does not take into account the correlated uncertainties among Gaia stars, including the systematic error of Gaia's wandering PM zero-point, which varies both as a function of sky position and field size \citep{Lindegren21a, Vasiliev21, warfield+2023a}. 

For their Gaia-based measurement of Draco~II, \cite{pace+2022} estimate the total size of this systematic error to be $0.022$~mas/yr using the relation given by \cite{Lindegren21a}, although they also note that the formula for this error given by \cite{Vasiliev21} suggests systematic errors up to 40\% larger. 
\cite{Battaglia+2021} additionally checks for a significant zero-point offset by measuring the weighted-average PMs of QSOs (as reported by Gaia) within a 7-degree radius around each galaxy. These offsets come out to $(\mu_{\alpha}^*, \mu_{\delta})_{\rm QSOs} = (0.001, 0.001)\pm 0.021$~mas/yr for Draco~II, leading to their conclusion that the statistical error dominates the uncertainty.
Therefore, our measurement made using only background galaxies provides a valuable, independent check on this assumption, being the only result that does not incorporate Gaia data.

\subsection{Methodological Systematic Uncertainty}
Comparing the results of \cite{pace+2022} and \cite{Battaglia+2021}, we also see that there can be a diversity of results among measurements starting from the same data set and using the same (or similar) methodology due to minor differences in choices/decisions made when implementing those methodologies. Differences between the measurements of \cite{pace+2022} and \cite{Battaglia+2021} can be attributed to how each team chose their sample of likely Draco~II stars and the slightly different set-ups of their GMMs. Where \cite{Battaglia+2021} only used Gaia~eDR3 astrometric and photometric data to determine galaxy membership, \cite{pace+2022} uses a combination of external photometric catalogs. The small differences in membership tagging is especially relevant for UFDs like Draco~II, where the number of sources available for measuring the PM is $\lesssim$~25.

Similarly, there are potential systematics, due to methodological choices, present in the HST-based results. For instance, we have already explored this in relation to the local- versus quadrant-corrections to the reflex PMs of background galaxies in \S\ref{ssec:gals}. There, we saw that the scale chosen at which to map the post-transformation residuals may affect whether we are under- or over-fitting for local residuals, but also may affect the measurement by changing the number of sources that are available to measure the PM overall.
Even though we include the scatter in the correction in the result's error budget, this extra systematic variance, if significant, may not be counted in full.
In order to address this issue for the \texttt{HubPUG} results, we have scaled the measured uncertainty on the weighted mean by a factor of $\sqrt{\chi^2_\nu} = \sqrt{\chi^2 / (N-1)}$, where $N$ is the number of Gaia stars or background galaxies used in the measurement. These rescaled uncertainties are what we present in Table~\ref{tab:pmresults}.

Comparing the \texttt{HubPUG} and \texttt{BP3M} results, we see that there is some level of offset despite starting with the same source catalogs. However, we can also recognize that, despite \emph{starting} with the same catalogs, the differing methodologies lead to source culling being performed at different steps in their procedures, which results in removing different sources (that are used in different parts of the calculation).

In another sense, the \texttt{BP3M} analysis presented here, using both the galaxies and stars, represents an entirely different treatment of the data. Rather than explicitly using the background galaxies to construct a reference frame, they are used as a way to deal with the systematic errors present in the Gaia-to-HST match (e.g. by having external-to-Gaia information about how the images should be aligned).
\texttt{BP3M}'s Bayesian approach to the frame transformation allows it to not (explicitly) "throw away" any Gaia star or background galaxy when tying each image to a sky frame. Draco~II stars are then cut based on their individual PM results when calculating the systemic PM. In effect, all cuts are then motivated by information internal to the calculation.
\texttt{HubPUG}, because it effectively does the calculation in reverse, requires pairing-down the Draco~II star list before the transformation (using external CMD-cuts), and then additional cuts are performed iteratively during the transformation. The Gaia star (or background galaxy) list is only touched when calculating the systemic PM.

\section{Conclusions} \label{sec:conclusions}

In this paper, we have presented multiple new measurements for the PM of Draco~II, all based on imaging from HST/ACS but utilizing three different methodologies and reference frames for the calculation.
We summarize each method as follows:
\begin{enumerate}
    \item \texttt{HubPUG} w/ Gaia stars (\S\ref{sec:method1})---Draco~II stars are used to construct the comoving frame, and the list of Draco~II stars is iteratively reduced to minimize the residuals between each image coordinate frame and the master coordinate frame. The PM of Draco~II is measured as the weighted mean of $N$ stars that trace the motion of the stationary Gaia frame versus the comoving frame.
    \item \texttt{HubPUG} w/ background galaxies (\S\ref{ssec:gals})---Draco~II stars are used to construct the comoving frame, and the list of Draco~II stars is iteratively reduced to minimize the residuals between each image coordinate frame and the master coordinate frame. The PM of Draco~II is measured as the weighted mean of $N$ background galaxies that trace the motion of a stationary frame constructed by background galaxies versus the comoving frame.
    \item \texttt{BP3M} (\S\ref{sec:bp3m})---Stars with Gaia positions and proper motions are used to simultaneously transform each \flc\ image to the Gaia sky frame, with background galaxies being used to control for systematics between the Gaia and HST frames by improving the internal alignment between HST images. The PM of Draco~II is measured as the correlated motion of individual Draco~II stars between epochs in this sky frame.
\end{enumerate}
Methods 2 and 3 also required us to implement additions/improvements into the associated machineries. For Method 2, we adapted the \texttt{HubPUG} software to be able to use background galaxies for the PM measurement, including correcting for HST detector-based systematics. This is the only measurement of Draco~II's PM not dependent on Gaia data. For Method 3, we used an adapted version of \texttt{BP3M} that implements two epochs of HST data, and is the only method that used Gaia positions and HST background galaxies simultaneously.

PM measurements for MW satellites are important for mapping the MW potential (e.g., \citealt{fritz2020}), measuring environmental effects on galaxies (e.g., \citealt{sacchi+2021}), and seeing the importance of the LMC's infall on the history of the MW system (e.g., \citealt{Patel2020}). UFDs are important tracers of these phenomena (e.g., there are many of them as an effect of the luminosity function of galaxies). However, UFD PMs are difficult to measure precisely because of their faintness and low star counts.
Two of our measurements---from Method~2 and Method~3---provide significant improvements on uncertainties over the measurements based on only Gaia data, with $\gtrsim1.5\times$ improvement. 
Both methods have uncertainties on their tangential velocities of $\lesssim5.0$~km/s. At this level, variance in orbit calculations will be overwhelmingly dominated by uncertainties on the Solar velocity vector, heliocentric distance, and the potential of the MW's halo (also shown by \citealt{pace+2022}). This indicates that both methods are robust for improving (or calculating at all) PM measurements in the low-source count UFD regime; i.e., measuring the PMs of UFDs to scientifically-useful precision beyond what would otherwise be possible with Gaia to the level that PM precision is no longer a limiting factor in studies. This will be valuable for future work. For instance, both \texttt{HubPUG} and \texttt{BP3M} should be able to measure precise PMs for the fourteen outer-halo UFDs targeted in HST Guest Observer Program 18068 (PI: N. Kallivayalil), which all have velocity uncertainties from Gaia $>50$~km/s.

As the uncertainties on PM measurements are inversely proportional to the time-baselines separating imaging epochs, maximizing these baselines will be crucial for making useful PM measurements of galaxies at distances $\gtrsim500$~kpc (e.g., M31 dwarf satellites), or for resolving internal motions (e.g., \citealt{libralato+2022}, \citealt{vitral+2025}). Because many of these galaxies were first observed with HST recently if at all (e.g., the HST M31 Treasury Program, executed 2019--2020; GO-15902, PI D. Weisz), and recognizing HST/ACS's limited lifetime, we will need to be able to make these precise astrometric measurements using observations from a combination of instruments moving forward. This includes combining archival HST/ACS, HST/WFC3, or HST/WFPC2 observations with those from JWST, ESA's Euclid, and/or the upcoming Nancy Grace Roman Space Telescope. Being able to combine these observatories for sub-pixel astrometry requires carefully considering and calibrating even the smallest cross-observatory systematic errors, taking into account our understanding of the PSF for different detectors as well as how each observatory's observational strengths and weaknesses affect the measurement (see, e.g., the work that was required to calibrate HST/WFPC2 astrometry to combine it with second-epoch HST/ACS data for PM measurements; \citealt{casettidinescu2023_starcenter}, \citealt{casettidinescu2024_starcenter}, \citealt{casettidinescu2024_and3}). In addition to the two epochs of HST/ACS data used in this paper, Draco~II has already been observed using JWST's NIRCam both in 2022 (PID 1334; PI D. Weisz) and 2024 (PID 4570; PI J. Warfield). Therefore, we plan to use the proper motion measurements presented here as calibration points for making these same measurements using JWST imaging as the second-epoch in future work.

\begin{acknowledgments}
This project was carried out in collaboration with the High-resolution Space Telescope PROper MOtion (HSTPROMO) Collaboration (\url{https://www. stsci.edu/\~marel/hstpromo.html}), a set of projects aimed at improving our dynamical understanding of stars, clusters, and galaxies in the nearby Universe through measurement and interpretation of PMs from HST, JWST, and Gaia. We thank the collaboration members for the sharing of their ideas and software.

This research is based on observations made with the NASA/ESA Hubble Space Telescope obtained from the Space Telescope Science Institute, which is operated by the Association of Universities for Research in Astronomy, Inc., under NASA contract NAS 5–26555. These observations are associated with programs GO-14734 (PI N. Kallivayalil) and GO-17473 (PI J. Warfield).

This work has also made use of data from the European Space Agency (ESA) mission Gaia (\url{https://www.cosmos.esa.int/gaia}), processed by the Gaia Data Processing and Analysis Consortium (DPAC, \url{https://www.cosmos.esa.int/web/gaia/dpac/consortium}). Funding for the DPAC has been provided by national institutions, in particular the institutions participating in the Gaia Multilateral Agreement.

Support for this work was provided by NASA through grants AR-15633, AR-16625, and GO-17473.
JTW acknowledges support from the Jefferson Scholars Foundation Graduate Fellowship.
KAM acknowledges supports from the University of Toronto’s Eric and Wendy Schmidt AI in Science Post-doctoral Fellowship, a program of Schmidt Sciences.
\end{acknowledgments}

%

\vspace{5mm}
\facilities{HST (ACS), Gaia, MAST}


\software{\texttt{HubPUG} \citep{warfield+2023a}, \texttt{BP3M} \citep{mckinnon+2024}, \texttt{GaiaHub} \citep{delPino+2022}, \texttt{hst1pass} \citep{hst1pass}, \texttt{DOLPHOT} \citep{dolphin2000b,dolphin2016}, \texttt{scikit-learn} \citep{scikit-learn}, \texttt{matplotlib} \citep{Hunter:2007}}



\bibliography{bibliography}{}
\bibliographystyle{aasjournal}



\end{document}